\documentclass[10pt]{iopart}
\usepackage[latin1]{inputenc}
\usepackage{ifthen,wasysym}

\newcounter{eqnumber}
\newcommand{\inlimine}[2]{\hfill\begin{minipage}[t]{7.5cm}
  \small\begin{flushleft}{#1}\end{flushleft}
  \ifthenelse{\equal{#2}{}}{\relax}{%
        \begin{flushright}
            \emph{#2}
        \end{flushright}}\end{minipage}\\[0.5cm]}
\newcommand{\D}{{\mathrm{d}}}
\newcommand{\E}{{\mathrm{e}}}
\newcommand{\average}[1]{\left<{#1}\right>}
\newcommand{\kt}{{k_\mathrm{B}T}}

\begin{document}

\title{%
Comment on: Inconsistency of the nonstandard definition of work}

\author{%
Luca Peliti}
\address{
Dipartimento di Scienze Fisiche,
Università ``Federico II''\\
Complesso Monte S. Angelo, 80126 Napoli (Italy)}

\date{%
October 8, 2007}

\begin{abstract}
    The objections raised by Vilar and Rubi [cond-mat arXiv:0707.3802v1]
    against the definition of the thermodynamical
    work appearing in Jarzynski's equality are shown to be
    misleading and inconsistent.
\end{abstract}

\inlimine{%
    Quid dices de primariis huius Gimnasii philosophis,
    qui aspidis pertinacia repleti, licet me ultro dedita
    opera millies offerente, nec Planetas, nec $\rightmoon$, nec
    perspicillum videre voluerunt? Verum ut ille aures,
    sic isti oculos, contra veritatis lucem obturarunt.}{%
        Galileo (in a letter to Kepler, 1610)}

\section{Introduction} In a recent post, Vilar and Rubi~(VR)~\cite{VR2} ascribe
to Imparato and Peliti~\cite{IP} the claim that the standard
definition of work
\begin{equation}\label{WFD:eq}
    \mbox{Work}=\mbox{Force}\times\mbox{Displacement},
\end{equation}
should be unconditionally replaced by a `nonstandard' definition
\begin{equation}\label{WIP:eq}
    \D W_\mathrm{IP}=-x\,\D f,
\end{equation}
in which the force and the displacement have their role
interchanged, when considering the work performed by a force on a
system. They argue that ``this `nonstandard' definition of work is
thermodynamically inconsistent at both the microscopic and
macroscopic scales and leads to non-physical results, including free
energy changes that depend on arbitrary parameters''. The dispute
arose from the claim set forth by Vilar and Rubi in a previous
post~\cite{VR1}, in which it was argued that the connection between
the microscopic work $W$ performed by a time-dependent force on a
system cannot be used to estimate free energy changes.

In the present note I shall argue the following points, which are
already clear to any honest reader of~ref.~\cite{IP}:
\begin{enumerate}
    \item The expression (\ref{WIP:eq}), surprising as it is, is
    a straightforward consequence of the standard definition of the
    \textit{thermodynamical} work performed on a system, for the
    special case considered in~\cite{IP}, namely, when a uniform
    but time-varying force is applied to a particle subject to a given
    potential;
    \item This expression yields, via a straightforward application
    of the First Principle of thermodynamics, a correct evaluation
    of the free-energy change of a thermodynamical system undergoing a
    reversible transformation;
    \item The `inconsistencies' claimed by VR to be produced by this
    expression of the work correspond to \textit{bona fide} energy
    differences which have observable consequences.
\end{enumerate}
I shall also argue that VR's confusions stem from the fact that the
thermodynamical work on a system represents the work done by the
system one considers on the external bodies which act on it, rather
than the work done on the system itself by the external bodies: a
point stressed, \textit{e.g.}, at the beginning of Gibbs's founding
book~\cite{Gibbs} on Statistical Mechanics, and that VR fail to
appreciate.

I shall first discuss these points in the context of equilibrium
thermodynamics. Further points are relevant when considering
manipulated systems, in particular small systems for which
fluctuations are important.

\section{Reversible work on the harmonic oscillator}
Let us consider a simple thermodynamical system,
\emph{i.e.}, a one-dimensional oscillator characterized by its mass
$m$ and spring constant $k$, kept at a fixed temperature $T$. The
system is described by the hamiltonian
\begin{equation}
    H(p,x)=\frac{p^2}{2 m}+\frac{1}{2}k x^2.
\end{equation}
In the following we shall focus only on the \emph{displacement}
degree of freedom, namely $x$. Its equilibrium distribution is given
by
\begin{equation}
    p^{\mathrm{eq}}(x)=\frac{\E^{-k x^2/2\kt}}{Z},
\end{equation}
where $Z$ is given by
\begin{equation}
    Z=\int \D x\;\E^{-k x^2/2\kt}=\sqrt{2\pi \kt/k}.
\end{equation}

We shall now apply a uniform, but time-varying, force
$f(t)$ to the system. We wish to evaluate the thermodynamical work
performed on it, as the applied force changes from $f_0=0$ to $f$,
so slowly, that \textit{the system can be considered to remain at
thermodynamical equilibrium at all times}. This is called the
\emph{reversible work} in thermodynamics.

Following the method described by J.~W.~Gibbs~\cite{Gibbs} and
R.~C.~Tolman~\cite{Tolman}, one proceeds as follows:
\begin{itemize}
    \item[1.] One writes down the hamiltonian of the system in the
    presence of the applied force:
    \begin{equation}\label{hamiltonian:eq}
        H(x,f)=\frac{1}{2}k x^2 -f \left(x-\gamma\right).
    \end{equation}
    Here $\gamma$ is defined as the point in which the potential of
    the applied force vanishes. This point might depend on $f$, but
    we shall momentarily assume that it is fixed. It is determined
    by the actual device used to apply the constant force on the
    system, as discussed in the following.
    \item[2.] One applies either Gibbs's
    equation~(117)~\cite[p.45]{Gibbs}, or Tolman's
    equation~(124.1)~\cite[p.542]{Tolman}, to obtain the
    thermodynamical work $\D W$ associated with a small variation $\D f$ of
    the applied force:
    \begin{equation}\label{dW:eq}
        \D W=\average{\frac{\partial H}{\partial f}}\,\D
        f=-\average{(x-\gamma)}\,\D f=-\left(\average{x}-\gamma\right)\,\D f.
    \end{equation}
    In this equation, $\average{A}$ is the canonical average of the
    function $A(x)$:
    \begin{equation}
        \average{A}=\frac{1}{Z}\int \D x\;A(x)\,\E^{-H(x,f)/\kt}.
    \end{equation}
    In our case, one obtains
    \begin{equation}\label{average:eq}
        \average{x}=\frac{f}{k},
    \end{equation}
    from which $\D W$ can be calculated via equation~(\ref{dW:eq}).
    \item[3.] One integrates the result with a variable force
    $f'$ from the initial value $f_0=0$
    to the final value $f$, obtaining
    \begin{equation}\label{deltaF:eq}
    \fl \qquad    \Delta F=\int_{0}^{f}\average{\left.\frac{\partial H}{\partial f}\right|_{f'}}\,\D
        f'=-\int_{0}^{f}\left(\frac{f'}{k}-\gamma\right)\,\D
        f'=-\frac{f^2}{2k}+\gamma f.
    \end{equation}
    In this expression, $\Delta F$ is the change in the Helmholtz
    free energy, $F=E-TS$. Since it is easy to see that in the
    present system the entropy $S$ does not change during the
    manipulation, we can equate it with the change in the
    \emph{internal} energy $E$. We have therefore
    \begin{equation}\label{deltaE:eq}
       \Delta E=-\frac{f^2}{2k}+\gamma f.
    \end{equation}
    \item[4.] Since the average value of the applied potential is given
    by
    \begin{displaymath}
        \average{U}=-f\left(\average{x}-\gamma\right),
    \end{displaymath}
    by subtracting it from the above result, we obtain the change of
    the energy of self-interaction of the spring
    \begin{equation}
        \Delta E^{\mathrm{int}}(f)=\Delta
        E-\average{U}=\frac{f^2}{2k}.
    \end{equation}
\end{itemize}

It would not be necessary to consider this elementary exercise in
statistical mechanics, were it not for the fact that in their recent
post J.~Vilar and M.~Rubi~\cite{VR2} (objecting to a similar
derivation contained in~\cite{IP}) have found that this result is
``inconsistent and unphysical both at the macroscopic and
microscopic level.'' Equation (\ref{dW:eq}) is the one that VR
incriminate. The two authors are chagrined by the following facts:
\begin{itemize}
    \item[1.] Let us first consider $\gamma=0$. Then the free-energy
    change~(\ref{deltaF:eq}) is negative. Now, non-spontaneous
    processes should lead to positive free-energy changes. This is in
    contrast with previous results, including ones on
    macromolecules~\cite{Liphardt}. Moreover
    this result holds for any system described by the hamiltonian
    (\ref{hamiltonian:eq}), including a macroscopic spring. This is
    in contrast with the results of elementary physics.
    \item[2.] Moreover, VR claim that the parameter $\gamma$ does
    not have any physical interpretation, and that therefore in this
    result the free-energy change does not depend on the actual physical system
    but rather on its mathematical description.
\end{itemize}

VR notice that the free-energy change given by equation
(\ref{deltaF:eq}) is negative since it also contains the potential
energy associated with the external force. They claim that it is
inconvenient to have the particular properties of the applied
external force embedded into the results. Therefore they deviate
from the definition of the thermodynamical work given by Gibbs, who
explicitly states~\cite[p.4, footnote]{Gibbs} that the energy
function of the statistical system should include ``that energy
which might be described as mutual to that system and external
bodies''. It lies on them, therefore, the burden to show that
\emph{their} `nonstandard' definition of the thermodynamical work is
preferable to Gibbs's and Tolman's one. They shun this burden by
failing to notice it.

They should however agree
that, if the potential energy of the interaction of the
system with the bodies that provide the constant force is taken into
account, both objections raised above disappear. VR proceed instead
as if the expression (\ref{deltaE:eq}) contained only the energy of
interaction of the system with itself.

I now show how the result (\ref{deltaE:eq}) corresponds to the
variation in the total energy (as defined in the above text by
Gibbs) when the external force is applied by two physically
reasonable devices. I shall then discuss why the apparent paradox of
point 1.\ is such only in the minds of the authors of
ref.~\cite{VR2} and their followers. But I now wish to stress the
point which probably lies at the heart of VR's confusion, by quoting
at length from Gibbs's treatise.
\setcounter{eqnumber}{\value{equation}}
\begin{quotation}
Returning to the case of the canonical distribution, we shall find
other analogies with thermodynamics systems, if we suppose, as in
the preceding chapters,\footnote{See especially Chapter~I, p.~4
(Note by JWG).} that the potential energy $(\epsilon_q)$ depends not
only upon the coordinates $q_1\ldots q_n$ which determine the
configuration of the system, but also upon certain cöordinates
$a_1$, $a_2$, etc. of bodies which we call \emph{external}, meaning
by this simply that they are not to be regarded as forming any part
of the system, although their positions affect the forces which act
on the system. The forces exerted by the system \textit{on these
bodies}\footnote{My italics (LP).} will be represented by
$-d\epsilon_q/da_1$, $-d\epsilon_q/da_2$, etc., while
$-d\epsilon_q/dq_1$\dots $-d\epsilon_q/dq_n$ represent all the
forces acting upon the bodies of the system, including those which
depend upon the position of the external bodies, as well as those
which depend only upon the configuration of the system itself. It
will be understood that $\epsilon_p$ depends only upon $q_1,\ldots
q_n,p_1,\ldots p_n$, in other words, that the kinetic energy of the
bodies which we call external forms no part of the kinetic energy of
the system. It follows that we may write \setcounter{equation}{103}
\begin{equation}
    \frac{d\epsilon}{da_1}=\frac{d\epsilon_q}{da_1}=-A_1,
\end{equation}
although a similar equation would not hold for differentiation
relative to the internal cöordinates.
\end{quotation}
\setcounter{equation}{\value{eqnumber}} Thus Gibbs's expression of
the elementary reversible work
\begin{equation}
    dW= -\sum_i \overline{A}_i\,da_i,
\end{equation}
(where, in Gibbs's notation, the bar denotes the average over a
canonical distribution) represents the average work \emph{done on
the external bodies by the system} (with changed sign), and
therefore, in particular, does not vanish even if the coordinates of
the system do not change over the time interval considered.

\subsection*{Electrostatic device}
To illustrate this point, let us set up a device for applying a
uniform but time-dependent force on our oscillator. We can use, for
instance, the following electrostatic device. Let us assume that the
mass of the oscillator carries a small charge $q$. We take two
point-like bodies at infinity, one with the charge $+Q$ and the
other with the charge $-Q$. We then let these two charged bodies
come closer and closer to the origin (the equilibrium point of the
oscillator), by letting the charge $+Q$ be situated at the point
$-X+\gamma$, and the charge $-Q$ at the point $X+\gamma$. Thus the
electric field acting on the oscillator at point $x$ is given by
\begin{eqnarray}\label{expansion:eq}
    E&=&\frac{Q}{4\pi\epsilon_0}\left[\frac{1}{(x-X-\gamma)^2}+
    \frac{1}{(x+X-\gamma)^2}\right]\nonumber\\
    &=& \frac{Q}{2\pi \epsilon_0}
    \frac{(x-\gamma)^2+X^2}{[(x-\gamma)^2+X^2]^2-4 (x-\gamma)^2
    X^2}\nonumber\\
    &\simeq &\frac{Q}{2\pi \epsilon_0}\left\{\frac{1}{X^2}
    +\frac{3 (x-\gamma)^2}{X^4}
    +\frac{5 (x-\gamma)^4}{X^6}+\cdots\right\}
\end{eqnarray}
If $X$ is large enough, then all terms beyond the first one are
negligible, for the expected excursions of the oscillator from the
origin. Then the force applied by the charge $Q$ is given by
\begin{equation}\label{force:eq}
    f=\frac{qQ}{2\pi\epsilon_0}\frac{1}{X^2}.
\end{equation}
Let us choose $Q$ such that, even for the largest force $f_1$ which
we wish to apply, $X$ is so large that the terms beyond the first in
equation (\ref{expansion:eq}) are negligible. Thus by moving the
charges $\pm Q$ from infinity to $\pm X+\gamma$, always
symmetrically around the point $\gamma$, we can apply a uniform but
time-varying force to our oscillator. It is now clear that $\gamma$,
far from being a fictitious parameter, corresponds to the location
of the center of the device by which a uniform force is applied to
the system we are studying. In order to change $\gamma$, external
work must be supplied to the apparatus.

Let us now evaluate the internal energy of the system as a function
of $X$. We have
\begin{eqnarray}
 \fl\qquad   E=\average{\frac{1}{2}k x^2+ U(x,X)}\nonumber\\
    \fl \qquad\quad {}=\frac{1}{Z}\int \D x\;
    \E^{-H(x,X)/\kt}\,\left[\frac{1}{2}k x^2
    +\frac{qQ}{4\pi\epsilon_0}\left(\frac{1}{x+X-\gamma}
    -\frac{1}{X-x+\gamma}\right)\right].
\end{eqnarray}
The first term yields
\begin{equation}
    \average{\frac{1}{2}k x^2}=\frac{1}{2}k
    \left[\average{(x-\average{x})^2}+\average{x}^2\right]=\frac{1}{2}\kt
    + \frac{1}{2}\frac{f^2}{k}.
\end{equation}
The first term is given by the equipartition theorem, and the second
by equation (\ref{average:eq}). One can expand the second term in
powers of $1/X$, obtaining
\begin{equation}
    \average{U(x,X)}=-\frac{qQ}{2\pi\epsilon_0}
    \left[\frac{1}{X^2}\average{(x-\gamma)}
    +\frac{1}{X^4}\average{(x-\gamma)^3}+\cdots\right].
\end{equation}
Thus, if $\average{(x-\gamma)^2}/X^2\ll 1$, we have
\begin{equation}
    \average{U(x,X)}=-\frac{qQ}{2\pi\epsilon_0}
    \frac{\average{(x-\gamma)}}{X^2}=-\frac{f^2}{k}+\gamma f,
\end{equation}
where we have exploited (\ref{force:eq}) and (\ref{average:eq}).
Summing up, we obtain
\begin{equation}
    E=\frac{1}{2}\kt -\frac{f^2}{2k}+\gamma f,
\end{equation}
in agreement with equation (\ref{deltaE:eq}).

\subsection*{Gravity-field device}
A simpler conceptual experiment can be set up imagining that the
oscillator mass is constrained to move along a line, which can be
rotated in the vertical plane. Let $m$ be the oscillator mass, $g$
the acceleration of gravity, and let the hinge be placed at
$x=\gamma$. If the line is now rotated clockwise by an angle
$\theta$, the oscillator mass will acted upon by a uniform force,
directed towards increasing values of $x$, and of intensity $mg \sin
\theta$. On the other hand, if the mass is at location $x$, its
height with respect to the horizontal line passing through the hinge
is given by $z=-(x-\gamma)\,\sin\theta$. It is then a simple matter
to evaluate the average of $U(x,\theta)$:
\begin{equation}
    \average{U(x,\theta)}=mg\average{z}=-mg \sin \theta
    \average{x-\gamma}=-\frac{f^2}{k}+\gamma f.
\end{equation}
Adding to it the average elastic energy $\frac{1}{2}k\average{x^2}$
we recover equation~(\ref{deltaE:eq}) again. But it is amusing to
verify that this result corresponds indeed to the work done by the
system on the external device. Let us consider the line to be tilted
by $\theta$, and the position of the oscillator to be $x$. Then the
oscillator applies to the rectilinear guide a torque
\begin{equation}
    \tau = m g \cos \theta \,(x-\gamma) .
\end{equation}
As the angle changes by $\D\theta$, this torque executes on the
guide a work \[\tau\,\D\theta= mg (x-\gamma)\,\D\sin\theta.\] The
\emph{reversible} elementary work made by the system on its
environment is given by the average of this expression, namely
\begin{equation}
    -\D W^{\mathrm{rev}}=\average{\tau}\,\D\theta=mg
    \left(\average{x}-\gamma\right) \,\D\sin\theta,
\end{equation}
where, according to equation~(\ref{average:eq}),
$\average{x}=f/k=mg\sin\theta/k$. The change in the internal energy
due to the transformation is given by $\D W^{\mathrm{rev}}$,
integrated between 0 and the final value of $\theta$. It is easy to
check that it yields again the result (\ref{deltaE:eq}).

When the rectilinear guide is tilted, the oscillator spring is
stretched and its elastic energy is increased. On the other hand,
the potential energy of the mass in the gravity field can either
increase or decrease, and the resulting total energy change can be
of either sign. If $\gamma=0$, one has, for instance
\begin{equation}
    \Delta E=-\frac{m^2 g^2 \sin^2\theta}{2k}=-\frac{f^2}{2k}.
\end{equation}
VR claim that this result is inconsistent, because a negative
free-energy change (which coincides in our case with the energy
change) would imply that the process is spontaneous, and that the
spring is unstable, in contradiction with elementary physics. They
fail to notice, however, that, \emph{if the rectilinear guide is
free to rotate around the origin}, the system is indeed unstable:
the guide would rotate till it reaches a vertical stand, with the
oscillator mass hanging on the spring. Thus, far from being
unphysical, the result yields the correct prediction for the
physical setup one is considering. Of course, in an actual
experiment, one would \emph{constrain} the guide at a given angle
$\theta$, and the oscillator will find equilibrium around a point
$\average{x}$ given by equation~(\ref{average:eq}).

\section{Reversible and fluctuating work}
The textbook definition of reversible work is the work performed
when the thermodynamic transformation is so slow that the system can
be considered to stay at thermodynamic equilibrium at all times. In
this case, the trajectory average coincides with the ensemble
average, at least if equilibrium statistical mechanics holds. Then
the performed work \emph{does not fluctuate}, and one trivially has
\begin{equation}
    W^{\mathrm{rev}}=\average{W^{\mathrm{rev}}}
    =-\kt \log \average{\E^{-W^{\mathrm{rev}}/\kt}}.
\end{equation}
VR claim that this equality does not hold, presumably because in
their mind the reversible work (which is a canonical average)
fluctuates. On the other hand, a clear distinction was made in
ref.~\cite{IP} between \emph{reversible} and \emph{fluctuating}
work, a distinction that VR chose to ignore. For the benefit of the
reader, I recall the definition of the infinitesimal fluctuating
work on a system whose microscopic state is denoted by $x=(x_i)$,
and described by the hamiltonian $H(x,\mu)$, depending on an
external parameter $\mu$:
\begin{equation}\label{fluctuating:eq}
    \D W=\frac{\partial H(x,\mu)}{\partial\mu}\,\D\mu.
\end{equation}
We then have, for a given infinitesimal change $\D\mu$,
\begin{equation}
    \D W^{\mathrm{rev}}=\average{\D W},
\end{equation}
where the average is taken with respect to the canonical
distribution with the given value of $\mu$. Notice that the
fluctuating work does not depend on the \emph{change} in the
microscopic state $x$ of the system, but on the \emph{change of the
external parameter} $\mu$, because it represents the work done by
the system on the external bodies that act on it. One can thus
understand why it does not vanish if $\mu$ is suddenly changed: if,
\emph{e.g.}, we suddenly push the charges $\pm Q$ closer to the
origin in the electrostatic device, we have to provide some work,
part of which changes the interaction energy of the oscillator with
the charges. By the same token, if we change $\gamma$,
\textit{e.g.}, by rigidly displacing the field-creating charges $Q$,
we have to provide work on the system, even if the oscillator's mass
does not move.

The distribution of the fluctuating work exhibits a number of
interesting properties, among which the remarkable equality
$\average{\E^{-W/\kt}}=\E^{-\Delta F/\kt}$ derived by
Jarzynski~\cite{Jarzynski}, and which Hummer and~Szabo~\cite{HS}
showed how to exploit in order to obtain information on the
\emph{equilibrium} free-energy landscape from nonequilibrium
experiments. VR object to this development, claiming that the above
definition of the fluctuating work is unphysical and inconsistent.
We have just seen how nicely it fits with equilibrium statistical
mechanics, as defined by Gibbs and explained by Tolman. However,
other quantities also exhibit remarkable distributions. Let us
consider a system described by the hamiltonian
\begin{equation}
    H(x,\mu)=H_0(x)-\mu F(x).
\end{equation}
Then the fluctuating work defined above is given by
\begin{equation}
    \D W =- F(x) \,\D\mu,
\end{equation}
and satisfies Jarzynski's equality. On the other hand, we can also
define the work $\D W_0$ by
\begin{equation}
    \D W_0=\mu \sum_i \frac{\partial F}{\partial x_i}\,\D{x}_i,
\end{equation}
which represents the work done by the environment on the system. As
recently discussed by Jarzynski~\cite{Jarz2} in more detail, this
work satisfies an identity found long ago by Bochkov
and~Kuzovlev~\cite{BK}, namely
\begin{equation}
    \average{\E^{-W_0/\kt}}=1.
\end{equation}
However, it is \emph{true} that it is difficult to exploit this
identity in order to recover information on an equilibrium quantity
like $\Delta F$. Indeed, what VR have brilliantly shown
in~\cite{VR1} is that $W_0$ cannot be used to reconstruct
free-energy landscape, but they fail to inform the reader
of~\cite{VR2} that their arguments concern the use of $W_0$, leaving
the impression that their objections concern $W$ and the use of the
Jarzynski equality. Now there is no problem in applying the
Jarzynski equality to $W$. The resulting $\Delta F$ contains a
contribution from the interaction between the system and the
environment which, contrary to VR's statements, is easily subtracted
off (see, e.g., the ``histogram method'' discussed
in~\cite{Seifert,IP2}). It is the responsibility of the researcher
to choose the most appropriate tools for one's task. One should
choose a spoon to eat one's soup and a spade to dig a hole: VR
appear to prescribe everybody to pick up the spoon and then they
lament that it is not possible to dig holes.

\section{Conclusions}
We have seen that VR's objections against ref.~\cite{IP} stem from a
biased and misleading reading of it, and from their failure to
appreciate some basic concepts in statistical mechanics. I am at a
loss to understand why as serious and competent physicists as VR
could fall in such blunders, unless their confusions arise from an
aprioristic hostility to the recent exciting developments in the
statistical mechanics of manipulated systems. In this case, they
would remind of Galileo's colleagues, cited in the letter I have
posted \emph{in limine}, who refused to look in the telescope
because it did not fit within their world view. If it is so, let
them be happy to encourage their followers to raise objections based
on even faultier arguments than their own~\cite{Perez}. I shall have
no more to say on their subject.

\ack I thank A. Imparato for many discussions and insights on this
matter.

\end{document}